\documentclass[12pt]{article}
\usepackage{amsmath}
\def\a{\alpha}
\def\b{\beta}

\def\t{\theta}

\def\be{\begin{equation}}
\def\ee{\end{equation}}
\def\arr{\begin{array}{rll}}
\def\ea{\end{array}}
\def\bea{\begin{eqnarray}}
\def\eea{\end{eqnarray}}

\def\N2{$N{=}2$}

\def\>{\rangle}
\def\<{\langle}
\def\+{\dagger}
\def\={\ =\ }

\begin{document}
\begin{titlepage}
\setcounter{page}{0}
\begin{flushright}
LMP-TPU--01/13  \\
\end{flushright}
\vskip 1cm
\begin{center}
{\LARGE\bf  Particle collisions on near horizon   }\\
\vskip 0.5cm
{\LARGE\bf extremal Kerr background }\\
\vskip 1cm
$
\textrm{\Large Anton Galajinsky\ }
$

\vskip 0.7cm
{\sl
Laboratory of Mathematical Physics, Tomsk Polytechnic University, 634050 Tomsk,\\
Lenin Ave. 30, Russian Federation} \\

{Email: galajin@tpu.ru}
\end{center}

\vskip 1cm
\begin{abstract}
\noindent
Recently, Ba\~nados, Silk and West analyzed a collision of two particles near the horizon of the extremal Kerr black hole and demonstrated that the energy in the center--of--mass frame can be arbitrarily large
provided the angular momentum of one of the colliding particles takes a special value.
As is known, the vicinity of the extremal Kerr black hole horizon can be viewed as a complete vacuum spacetime
in its own right. In this work, we consider a collision of two neutral particles within the context of the near horizon extremal Kerr geometry and demonstrate that the energy in the center--of--mass frame is finite for any admissible value of the particle parameters. An explanation of why the two approaches disagree on the Ba\~nados--Silk--West effect is given.

\end{abstract}

\vspace{0.5cm}

PACS: 04.70.Bw\\ \indent
Keywords: near horizon black holes, Ba\~nados--Silk--West effect
\end{titlepage}

\renewcommand{\thefootnote}{\arabic{footnote}}
\setcounter{footnote}0

\noindent
{\bf 1. Introduction}

\vskip 0.5cm

In a recent work \cite{bsw}, Ba\~nados, Silk and West (BSW) considered a collision of two particles near the horizon of the extremal Kerr black hole and demonstrated that the energy in the center--of--mass frame can be arbitrarily large provided the angular momentum of one of the colliding particles takes a special value. This work stimulated an intensive debate \cite{BCG}--\cite{MW} which continues today. Various generalizations of the BSW effect were studied in the literature. These include nonextremal black holes, charged black holes, black holes surrounded by matter, consideration of nonequatorial geodesic particles etc.

In the original work \cite{bsw}, two particles of equal mass and equal energy falling freely from rest at infinity and approaching the extremal Kerr black hole on the equatorial plane were considered.
The energy in the center--of--mass frame was computed with the use of the first integrals of the geodesic equations.
Evaluating the energy in the limit in which the radial coordinate tends to the horizon radius, the critical value of
the particle angular momentum at which the energy blows up was determined. Such an ultrahigh energy collision is known in the literature as the BSW effect.

It has been known for some time that the vicinity of the extremal Kerr black hole horizon can be viewed as a complete vacuum spacetime
in its own right \cite{bh} (see also a related earlier work \cite{zas}). In particular, the near horizon extremal Kerr (NHEK) geometry is a cornerstone of the Kerr/CFT--correspondence \cite{str} which is being extensively investigated nowadays (for a review see \cite{com}). It is then important to confront the NHEK setup with the BSW analysis and to investigate whether the ultrahigh energy collisions take place within the context of the NHEK geometry.

In this paper we consider this issue and demonstrate that the NHEK geometry does not support the ultrahigh energy collision. Note that our analysis does not imply that the BSW result is incorrect. The original Kerr black hole and the NHEK geometry differ in many respects. To mention a few, the latter is not asymptotically flat,
it has a larger symmetry group, in case of a vanishing electric charge the NHEK metric involves the physical parameter as an overall factor only, the second rank Killing tensor is reducible etc. The goal of this paper is to demonstrate that the two geometries show different attitudes to the ultrahigh energy collision and to provide a possible explanation.

The work is organized as follows. In Sect. 2 we recapitulate the BSW results in a slightly more general framework of the near horizon extremal Kerr--Newman geometry. The near horizon Kerr geometry and the near horizon Reissner--Nordstr\"om geometry can be retrieved as particular cases. In Section 3 a similar analysis is carried out within the context of the NHEK geometry. In particular, we study in detail the radial equation of motion and the forward in time condition and establish the range of parameters for which a particle can reach the horizon. The energy in the center--of--mass frame is computed in the near horizon limit and is shown to be finite for any admissible value of the particle parameters. In the concluding Section 4 we confront the two pictures. It is shown that, while the particle angular momenta can be consistently identified in both the frameworks, a link between the particle energies is problematic. An explanation of why the two approaches disagree on the BSW effect is given.

\vskip 0.5cm

\noindent
{\bf 2.  BSW effect}

\vskip 0.5cm

In Boyer--Lindquist--type coordinates the Kerr--Newman solution of the Einstein-Maxwell equations reads\footnote{We use the metric with mostly minus signature and the relativistic units in which $c=1$, $G=1$. In these conventions the Einstein--Maxwell equations read
$R_{nm}-\frac 12 g_{nm} R=-2(F_{ns} {F_m}^s-\frac 14 g_{nm} F^2)$, $\partial_n (\sqrt{-g} F^{nm})=0$.}
\bea\label{kna}
&&
ds^2=\frac{\Delta}{\rho^2} \left(d t-a \sin^2 \theta d \phi\right)^2
-\frac{\rho^2}{\Delta} d r^2 - \rho^2 d\theta^2
-\frac{\sin^2\theta}{\rho^2}\left(a d t-(r^2+a^2) d \phi\right)^2,
\nonumber\\[2pt]
&&
A=\frac{q r}{\rho^2}\left( dt - a\sin^2\t d \phi \right), \quad \Delta=r^2+a^2-2Mr+q^2, \quad
\rho^2=r^2+a^2\cos^2\theta.
\eea
The parameters $M$, $a$ and $q$ are linked to the mass, angular momentum and electric charge of
the black hole, respectively.
In what follows we discuss only the extremal solution, which occurs if $\Delta$ has a double zero at the horizon radius $r_{h}=M=\sqrt{a^2+q^2}$.
The Kerr--Newman solution (\ref{kna}) is invariant under the time translation and rotation around the symmetry axis
\be\label{tp}
t'=t+\a, \qquad
\phi'=\phi+\b.
\ee
It also possesses a hidden symmetry described by the second rank Killing tensor \cite{car}.

Before we proceed to discuss the BSW effect, let us make a comment on the gauge choice.
The equations of motion describing a massive charged particle coupled to the Einstein--Maxwell background read
\be\label{AC}
m \left(\frac{d^2 x^n}{d s^2}+\Gamma^n_{mp} \frac{d x^m}{d s}\frac{d x^p}{d s}  \right)=e g^{nm} F_{mp}
\frac{d x^p}{d s},
\ee
where $m$ and $e$ are the mass and the electric charge of the particle, respectively, $F_{mp}=\partial_m A_p-\partial_p A_m$ is the electromagnetic tensor and $\Gamma^n_{mp}$ are the Christoffel symbols.
In general, the coordinate transformation $x'^n=x^n+\xi^n(x)$ generated
by a Killing vector field with components $\xi^n(x)$ leaves the background vector potential invariant provided
\be\label{sc}
\xi^m F_{mn}+\partial_n (\xi^m A_m)=0.
\ee
In view of (\ref{AC}) and (\ref{sc})
each Killing vector field gives rise to the integral of motion
\be\label{ekv}
\xi^n (x) \left(m g_{nm} \frac{d x^m}{d s}+e A_n\right).
\ee

The Einstein-Maxwell equations are invariant under the gauge transformation $A'_n(x)=A_n(x)+\partial_n \lambda(x)$. Taking into account (\ref{sc}), one concludes that a gauge transformation preserves symmetries of a gauge field one--form provided
\be
\xi^n \partial_n \lambda=\mbox{const}.
\ee
In view of (\ref{tp}), for the Kerr--Newman solution the arbitrariness in defining a gauge field one--form amounts to adding a constant contribution to the gauge potential. In general, this constant is set to zero by imposing a boundary condition that the gauge potential vanishes at spatial infinity. However, in the context of the near horizon geometry which we discuss below such a boundary condition is no longer valid. In particular, the near horizon gauge potential is linear in the radial coordinate. Because the gauge transformation shifts the value of the first integral (\ref{ekv}) by the constant $e\xi^n \partial_n \lambda$ and the energy in the center--of--mass frame nonlinearly depends on the first integrals, in what follows we consider only neutral particles and set $e=0$ throughout.

Let us consider a collision of two neutral particles of equal mass propagating on the Kerr--Newman background.
Focusing on the equatorial plane and taking into account the invariance of the background fields under the time translations and rotations around the symmetry axis, from the first integrals
\be\label{first}
E=m g_{tn} \frac{d x^n}{d s}, \qquad L=m g_{\phi n} \frac{d x^n}{d s}, \qquad g_{nm} (x) \frac{d x^n}{d s} \frac{d x^m}{d s}=1,
\ee
where $E$ is the energy of a particle and $L$ is the component of momentum parallel to the symmetry axis, one finds
\bea\label{dr}
&&
\frac{dt}{ds}=\frac{1}{r^2 \Delta} \left(a(2 M r-q^2) (a \mathcal{E}+l)+r^2 (r^2+a^2) \mathcal{E} \right), \quad
\nonumber\\[2pt]
&&
\frac{d\phi}{ds}=\frac{1}{r^2 \Delta} \left((2 M r-q^2) (a \mathcal{E}+l)-r^2 l\right),
\nonumber\\[2pt]
&&
\frac{dr}{ds}=\pm \frac{1}{r^2}\sqrt{r^2 (r^2+a^2) {\mathcal{E}}^2+(2 M r-q^2) {(a \mathcal{E}+l)}^2 -r^2 l^2-r^2 \Delta},
\eea
where $\mathcal{E}=E/m$, $l=L/m$. Computing the energy for a pair of particles in the
center--of--mass frame
\be\label{cm}
E_{c.m.}=m \sqrt{2} \sqrt{1+g_{mn} \frac{d x_1^m}{d s} \frac{d x_2^n}{d s}}
\ee
and taking the near horizon limit $r \to M$, one gets (see also \cite{KerrN})
\bea\label{E}
&&
E_{c.m.} (r \to M)=
\\[2pt]
&&
\quad \quad =m\sqrt{\frac{ {(\mathcal{E}_1+\mathcal{E}_2)}^2}{ \mathcal{E}_1 \mathcal{E}_2}+\frac{(l_1 \mathcal{E}_2- l_2 \mathcal{E}_1)(a {\mathcal{E}_1}^2+l_1 \mathcal{E}_1-a)}{\mathcal{E}_1 (\mathcal{E}_1 (q^2+2 a^2)+a l_1) } -\frac{(l_1 \mathcal{E}_2- l_2 \mathcal{E}_1)(a {\mathcal{E}_2}^2+l_2 \mathcal{E}_2-a)}{\mathcal{E}_2 (\mathcal{E}_2 (q^2+2 a^2)+a l_2) }   }.
\nonumber
\eea
In particular, setting $\mathcal{E}_1=\mathcal{E}_2=a=1$ and $q=0$, one reproduces the result in \cite{bsw} for the extremal Kerr black hole\footnote{Our convention for $l$ differs from that in \cite{bsw} by the flip of sign.}
\be\label{energy}
E_{c.m.} (r \to 1)=m \sqrt{2} \sqrt{\frac{l_2+2}{l_1+2}+\frac{l_1+2}{l_2+2}}.
\ee
Thus, the BSW effect occurs for the critical value $l_{1,2}=-2$, at which the energy (\ref{energy}) blows up.

Note that within the BSW framework (in which $M=a=1$, $q=0$) a particle which is originally at rest at infinity ($\mathcal{E}=1$) is governed by the effective potential (see Eq. (\ref{dr}) above)
\be
V(r)=1-\frac{2}{r}+\frac{l^2}{r^2}-\frac{2{(l+1)}^2}{r^3}
\ee
and can reach the horizon provided the following restriction on the angular momentum:~\footnote{Recall that our convention for $l$ differs from that in \cite{bsw} by the flip of sign.}
\be
-2\leq l \leq 2(1+\sqrt{2})
\ee
holds \cite{bsw}.

\vskip 0.5cm

\noindent
{\bf 3. Particle collisions on NHEK background}

\vskip 0.5cm

It is then interesting to carry out a similar analysis within the NHEK setup \cite{bh}.
In order to derive the NHEK geometry, one first redefines the coordinates
\be\label{ct}
r \quad \rightarrow \quad M + \epsilon r_0 r, \qquad t \quad \rightarrow \quad \frac{t r_0}{\epsilon}, \qquad
\phi \quad \rightarrow \quad \phi+
\frac{t a }{\epsilon r_0},
\ee
where $r_0^2=M^2+a^2$, in
such a way that in the new coordinate system the horizon is located at $r=0$.
Then one takes the limit $\epsilon \rightarrow 0$ which yields \footnote{
Note that in modern literature on the Kerr/CFT--correspondence (see, e.g., Ref. \cite{Hart} and references therein) its is customary to implement the near horizon limit so as to directly produce the conventional ${AdS}_2$ metric $r^2dt^2-\frac{dr^2}{r^2}$ as a part of the near horizon geometry. This implies that $t$ in Eq. (\ref{nhm}) has the dimension ${[\mbox{mass}]}^{-1}$. In this convention $\mathcal{E}$ and $l$ which enter Eq. (\ref{dr1}) have the dimensions ${[\mbox{mass}]}^2$ and $[\mbox{mass}]$, respectively. An alternative possibility which fits to the conventional dimension counting is to consider $r \to M + \epsilon r$, $t \to \frac{t}{\epsilon}$, $\phi \to \phi+
\frac{t a }{\epsilon (M^2+a^2)}$ instead of (\ref{ct}) which yield the metric (\ref{nhm}) with $t$ changed by $\frac{t}{M^2+a^2}$ \cite{bh}. The former and the latter pictures are thus related by the rescalings
$t \to \frac{t}{M^2+a^2}$,
$\mathcal{E}\to (M^2+a^2) \mathcal{E}$, with $l$ being unchanged. Worth mentioning also is that the near horizon vector potential is usually derived from the near horizon field strength two--from. For more details on the derivation of the near horizon black hole geometries in diverse dimensions and further references to the original literature see, e.g., a recent work \cite{g3}. }
\bea\label{nhm}
&&
ds^2=\rho_0^2 \left(
r^2dt^2-\frac{dr^2}{r^2}
-d\theta^2 \right) - \frac{{(M^2+a^2)}^2\sin^2\theta}{\rho_0^2} \left(d\phi+\frac{2 a M}{M^2+a^2} r dt\right)^2,
\\[2pt]
&&
A=\frac{q}{\rho_0^2} \left((M^2-a^2 \cos^2{\theta}) r dt+ a M \sin^2{\theta} d\phi\right), \qquad \quad \rho_0^2=M^2+a^2\cos^2\theta.
\nonumber
\eea
This is a vacuum solution of the Einstein-Maxwell equations \cite{bh}.

Focusing on the equatorial plane, from Eq. (\ref{nhm}) and the first integrals which have the same structure as in (\ref{first}) one derives
\bea\label{dr1}
&&
\frac{dt}{ds}=\frac{\mathcal{E}}{M^2 r^2}-\frac{2 a l}{M (M^2+a^2) r}, \qquad \frac{d\phi}{ds}=-\frac{(M^2-4 a^2) l}{{(M^2+a^2)}^2}
-\frac{2 a \mathcal{E}}{M (M^2+a^2) r},
\nonumber\\[2pt]
&&
\frac{dr}{ds}=\pm \frac{1}{M}\sqrt{\frac{{\mathcal{E}}^2}{M^2}-\frac{4 a l \mathcal{E} }{M (M^2+a^2)} r-\frac{[l^2 (M^2-4 a^2)+{(M^2+a^2)}^2] }{{(M^2+a^2)}^2} r^2}.
\eea
Computing the energy for a pair of particles in the
center--of--mass frame and taking the near horizon limit $r\to 0$, one gets
\be\label{EE}
E_{c.m.} (r \to 0)=m \sqrt{\frac{1}{\mathcal{E}_1 \mathcal{E}_2} \left( {(\mathcal{E}_1+\mathcal{E}_2)}^2+\frac{(q^2+a^2)}{{(q^2+2a^2)}^2} {(l_1 \mathcal{E}_2-l_2 \mathcal{E}_1)}^2\right)},
\ee
which is finite for any value of particle parameters.~\footnote{Below we demonstrate that the option $\mathcal{E}=0$ for one of the colliding particles is ruled out as unphysical.}

Let us discuss admissible values of particle parameters in more detail. These are determined by the effective potential $V(r)$ in the radial equation for an ingoing particle \footnote{Here and in what follows we discuss the extremal Kerr black hole only and set $M=a=1$, $q=0$.}
\be\label{req}
\frac{dr}{ds}=-\sqrt{\mathcal{E}^2-V(r)},
\ee
the forward in time condition $\frac{dt}{ds}>0$,
and the requirement that the particle can reach the horizon. From (\ref{dr1}) one finds
\be\label{EP}
V_{NH} (r)=2l\mathcal{E}r+\left(1-\frac{3 l^2 }{4} \right) r^2.
\ee

For $\left(1-\frac{3 l^2 }{4} \right)>0$ the potential is a parabola, which opens upward
\be
V_1={\left(r \sqrt{1-\frac{3 l^2}{4}} +\frac{l \mathcal{E}}{\sqrt{1-\frac{3 l^2}{4}}} \right)}^2-{\left(\frac{l \mathcal{E}}{\sqrt{1-\frac{3 l^2}{4}}} \right)}^2.
\ee
The cases $l \mathcal{E}>0$ and $l \mathcal{E}<0$ prove to be qualitatively similar and the solution of the radial equation (\ref{req}) can be written in a uniform way
\be\label{r}
r(s)=\frac{1}{1-\frac{3 l^2}{4}}  \left[ |\mathcal{E}| \sqrt{1+\frac{l^2}{4}} \cos{\left((s-s_0) \sqrt{1-\frac{3 l^2}{4}}  \right)}-l \mathcal{E} \right],
\ee
where $s_0$ is a constant of integration and it is assumed that $(s-s_0) \sqrt{1-\frac{3 l^2}{4}} \in [0,\pi]$. Choosing for definiteness $s_0=0$, one can readily verify that
the particle starts at $r(0)=\frac{|\mathcal{E}|}{\sqrt{1+\frac{l^2}{4}}+\frac{l \mathcal{E}}{|\mathcal{E}|}}$ and reaches the horizon located at $r=0$ in finite proper time $\tilde s=\frac{1}{\sqrt{1-\frac{3 l^2}{4}}} \arccos{\left( \frac{l \mathcal{E}}{|\mathcal{E}|\sqrt{1+\frac{l^2}{4}}} \right)}$. Substituting (\ref{r}) in
the leftmost equation entering the first line in Eq. (\ref{dr1}), one can verify that the forward in time condition $\frac{dt}{ds}>0$ selects $\mathcal{E}>0$, while $l$ is allowed to be positive, negative or zero.

For $\left(1-\frac{3 l^2 }{4} \right)<0$ the potential is a parabola, which opens downward
\be
V_2=-{\left(r \sqrt{\frac{3 l^2}{4}-1} -\frac{l \mathcal{E}}{\sqrt{\frac{3 l^2}{4}-1}} \right)}^2+{\left(\frac{l \mathcal{E}}{\sqrt{\frac{3 l^2}{4}-1}} \right)}^2.
\ee

If $l \mathcal{E}>0$ the vertex of the parabola lies in the first quadrant and the range of $r$ which admits the fall to the horizon reads $r<\frac{l \mathcal{E}}{\frac{3 l^2}{4}-1}$. The solution of the radial equation (\ref{req}) has the form
\be\label{r2}
r(s)=\frac{|\mathcal{E}|}{\frac{3 l^2}{4}-1} \left(|l|-\sqrt{\frac{l^2}{4}+1} \left( \frac{1+p^2}{2p} \right) \right), \qquad p=e^{-(s-s_0) \sqrt{\frac{3 l^2}{4}-1}},
\ee
where $s_0$ is a constant of integration, and it is assumed that $(s-s_0) \in [0,\infty)$. Choosing for definiteness $s_0=0$, one can check that
the particle starts at $r(0)=\frac{|\mathcal{E}|}{|l|+\sqrt{\frac{l^2}{4}+1}}$ and reaches the horizon in finite proper time $\tilde s=-\frac{1}{\sqrt{\frac{3 l^2}{4}-1}} \ln{\left( \frac{|l|-\sqrt{\frac{3 l^2}{4}-1}}{\sqrt{\frac{l^2}{4}+1}}\right)}$. With the use of (\ref{r2}) one can further verify that the forward in time condition $\frac{dt}{ds}>0$ selects $\mathcal{E}>0$ and $l>0$.

If $l \mathcal{E}<0$ the vertex of the parabola lies in the second quadrant and the potential is the descending leg which starts at $r=0$. Because the potential is unbounded below,
${(\frac{dr}{ds})}^2$ may grow unbounded. Such potentials are conventionally ruled out as unphysical. By the same reason the value $\mathcal{E}=0$  advocated recently in \cite{Z2} is unacceptable.

The instance $\left(1-\frac{3 l^2 }{4} \right)=0$ can be considered likewise. The potential is a linear function of $r$ with $\mathcal{E}>0$ and $l=\frac{2}{\sqrt{3}}$.

To summarize, for an ingoing particle moving on the NHEK background the admissible values of the parameters are $\mathcal{E}>0$ and $l>-\frac{2}{\sqrt{3}}$.

\vskip 0.5cm

\noindent
{\bf 4. Discussion}

\vskip 0.5cm

A few comments are in order. First of all, it should be understood that the results obtained within the conventional Kerr geometry and the NHEK setup can not be related by a
coordinate transformation. It is true that the derivation of the NHEK metric involves the coordinate transformation (\ref{ct}). However, it becomes singular in the near horizon limit $\epsilon \to 0$.
The NHEK geometry differs from the original Kerr background in many respects. The spacetime is not asymptotically flat. In the near horizon limit the $U(1)\times U(1)$--isometry of the Kerr metric is extended to
$SO(2,1)\times U(1)$ which in addition to (\ref{tp}) involves dilatations and special conformal transformations \cite{bh}. The larger isometry renders the near horizon Killing tensor reducible \cite{g2,go}.
Most notably, if the charge $q$ is zero the NHEK metric involves the physical parameter as an overall factor only.

As we have seen above, admissible values of particle parameters differ in the two pictures as well. In order to comprehend whether they actually have the same meaning and can be compared, let us consider the integrals of motion (\ref{first}) in the first picture
\be
\mathcal{E} ds=g_{tt} dt+g_{t \phi} d \phi, \qquad l ds=g_{\phi t} dt+g_{\phi \phi} d \phi
\ee
and apply to them the coordinate transformation (\ref{ct}) followed by the near horizon limit $\epsilon \to 0$. In order to avoid confusion, in what follows we use tilde to designate coordinates and particle parameters within the NHEK framework. Setting the black hole charge $q$ to zero and taking into account that in this case $M^2=a^2$ and $r_0^2=2 M^2$, one finds
\bea\label{nhl}
&&
\lim_{\epsilon \to 0} l ds=l d\tilde s=\lim_{\epsilon \to 0} (g_{\phi t} dt+g_{\phi \phi} d \phi)=-\frac{4 a M^2 \sin^2{\tilde\theta}}{\rho_0^2} (a d \tilde\phi+M \tilde r d \tilde t),
\nonumber\\[2pt]
&&
\lim_{\epsilon \to 0} \mathcal{E} ds=\mathcal{E} d \tilde s=\lim_{\epsilon \to 0} (g_{t t} dt+g_{t \phi} d \phi)=\frac{2 M^2 \sin^2{\tilde\theta}}{\rho_0^2} (a d \tilde\phi+M \tilde r d \tilde t),
\eea
where $\rho_0^2=M^2(1+\cos^2{\tilde\theta})$. It is noteworthy that these two expressions differ only by a constant factor.
Evaluating the explicit form of the first integrals for a particle propagating on the NHEK background
\bea
&&
\tilde l d \tilde s=g_{\tilde \phi \tilde t} d \tilde t+g_{\tilde\phi \tilde\phi} d \tilde\phi=
-\frac{4 a M^2 \sin^2{\tilde\theta}}{\rho_0^2} (a d \tilde\phi+M \tilde r d \tilde t),
\nonumber\\[2pt]
&&
\mathcal{\tilde E} d \tilde s=g_{\tilde t \tilde t} d \tilde t+g_{\tilde t \tilde\phi} d \tilde\phi=\rho_0^2 {\tilde r}^2 d \tilde t-\frac{4 M^3  \sin^2{\tilde\theta} \tilde r}{\rho_0^2} (a d \tilde\phi+M \tilde r d \tilde t),
\eea
one concludes that, while the angular momenta in both the pictures can be identified $l=\tilde l$,
the energy parameters $\mathcal{E}$ and  $\mathcal{\tilde E}$ are not related. Their magnitudes thus have different meanings and are not directly comparable.

Yet, an amazing feature of the near horizon limit (\ref{nhl}) is that it implies
\be
l=-2 a \mathcal{E}
\ee
which is precisely the BSW critical value (cf. Eq. (\ref{E}) above)! The last observation makes it clear why the two approaches disagree on the BSW effect.
In the first approach, one calculates the energy in the center--of--mass frame and then takes the near horizon limit in that expression. In the second approach, one first takes the near horizon limit for the metric and then computes the energy for a pair of particles in the center--of--mass frame in the new geometrical setup. Thus, not only the energy parameters of individual particles are not directly comparable in the two pictures but the quantities (\ref{E}) and (\ref{EE}) have different meanings as well. In fact, it is straightforward to verify that the coordinate transformation (\ref{ct}) applied to (\ref{cm}) and followed by $\epsilon \to 0$ yields a divergent expression even before fixing any critical value for the particle parameters.

\newpage

\noindent{\bf Acknowledgments}\\

\noindent
We thank S.T. McWilliams and O.B. Zaslavskii for useful comments.
This work was supported by RF Federal Program "Kadry" under the contract No 16.740.11.0469,
MSE Program "Nauka" under the grant No 1.604.2011 and by the RFBR grant No 13-02-90602-Arm.
{
\end{document}